# Antiresonance-Like Behavior in Carrier-Envelope-Phase-Sensitive Optical-Field Photoemission from Plasmonic Nanoantennas

*P. D. Keathley[1,‡,\*], W. P. Putnam[1,2,3,‡], P. Vasireddy[1], R. G. Hobbs[1,4], Y. Yang[1], K. K. Berggren[1], and F. X. Kärtner[1,2,5]*

[1]Department of Electrical Engineering and Computer Science and Research Laboratory of Electronics,
Massachusetts Institute of Technology, 77 Massachusetts Ave., Cambridge, MA 02139, USA
[2]Department of Physics and Center for Ultrafast Imaging, University of Hamburg, Luruper Chaussee 149, 22761 Hamburg, Germany
[3]Northrop Grumman Corporation, NG Next, 1 Space Park Blvd., Redondo Beach, CA 90278, USA
[4]Centre for Research on Adaptive Nanostructures and Nanodevices (CRANN), Advanced Materials and Bio-Engineering Research Centre (AMBER), and School of Chemistry, Trinity College Dublin, Dublin 2, Ireland
[5]Center for Free-Electron Laser Science and Deutsches Elektronen-Synchrotron (DESY), Notkestraße 85, 22607 Hamburg, Germany
[‡]These authors contributed equally to this work.
\*Corresponding author: pdkeat2@mit.edu

Optical-field-driven photoemission occurs when the electric field of an optical pulse bends the potential barrier of a material surface such that electron tunneling occurs before the field reverses polarity. At the surface of nanoscale structures, such as ultra-sharp nanoscale tips or metallic nanoantennas, the electric fields of ultrafast optical pulses are strongly enhanced. These enhanced fields can drive optical-field photoemission (*i.e.* optical-tunneling) and thereby generate and control electrical currents at frequencies exceeding 100 THz [1–11]. A hallmark of such optical-field photoemission is sensitivity of the total emitted current to the carrier-envelope phase (CEP) of the driving optical pulse, and this CEP-sensitivity has been studied using both atomic and solid-state systems[1–3,7,11–17]. Given the quasi-static nature of optical-field emission and the nontrivial dependence of the emission rate on the instantaneous electric field strength, the CEP-sensitive component of the emitted photocurrent is highly sensitive to the energy of the optical pulse, and should carry information about the underlying sub-cycle dynamics of electron emission. Here we examine CEP-sensitive photoemission from plasmonic gold nanoantennas excited with few-cycle optical pulses of increasing energy. We observe antiresonance-like features in the CEP-sensitive photocurrent; specifically, at a critical pulse energy, we observe a sharp dip in the magnitude of the CEP-sensitive photocurrent accompanied by a sudden shift of $\pi$-radians in the phase of the photocurrent. Using a quasi-static tunneling emission model, we find that these antiresonance-like features arise due to competition between electron emission from neighboring optical half-cycles, and that they are highly sensitive to the precise shape of the driving optical waveform at the surface of the emitter. As the underlying mechanisms that produce the antiresonance-like features are a general consequence of nonlinear, field-driven photoemission, the antiresonance-like features could be used to probe sub-optical-cycle, sub-femtosecond emission processes, not only from solid-state emitters, but also from gas-phase atoms and molecules. Beyond applications in the study of ultrafast, field-driven electron physics, an understanding of these antiresonance-like features will be critical to the development of novel photocathodes for future time-domain metrology and microscopy applications that demand both attosecond temporal and nanometer spatial resolution.

Our basic experimental setup is shown in Fig. 1a. We illuminate triangular, gold nanoantennas (nano-triangles) with few-cycle pulses of near-infrared light. Pulses arrive at a repetition rate of 78.4 MHz and have a central wavelength of 1177 nm and a duration of ≈ 2.5 optical cycles (≈ 10 fs). By aligning the polarization of the incident light along the length of the triangular nanoantennas, field-enhancement factors of ≈ 30 × can be achieved near the apices of the triangles due to both plasmonic resonance as well as geometric enhancement effects[1,9,10]. The triangular shape of the antennas was chosen to break the system's inversion symmetry, thereby increasing the CEP-sensitivity of the emitted photocurrent when driven into the optical-field photoemission regime[1]. The gold nanoantennas sit on a conductive layer of indium tin oxide (ITO), which behaves similarly to a dielectric at optical wavelengths, but remains conductive at low frequencies thus enabling charge extraction. When illuminated, the triangular nanoantennas predominantly emit electrons from their apices due to the enhanced peak surface field there. Following emission, the electrons are swept away through the interstitial air and across an insulating gap by a DC bias where they are collected by another ITO layer. The total photocurrent is detected using a transimpedance amplifier. An optical micrograph of our device and a scanning electron micrograph (SEMs) of our gold nanoantennas are shown in Fig. 1b. We should note that our experimental geometry closely parallels that described in Ref.[1], and our device closely resembles an optically-driven vacuum tube: optical pulses excite electrons from our nanoantenna-based photocathode (labeled "emitter" in Fig. 1b), and these emitted electrons are swept across a gap to a positively biased anode (labeled "collector" in Fig. 1b).

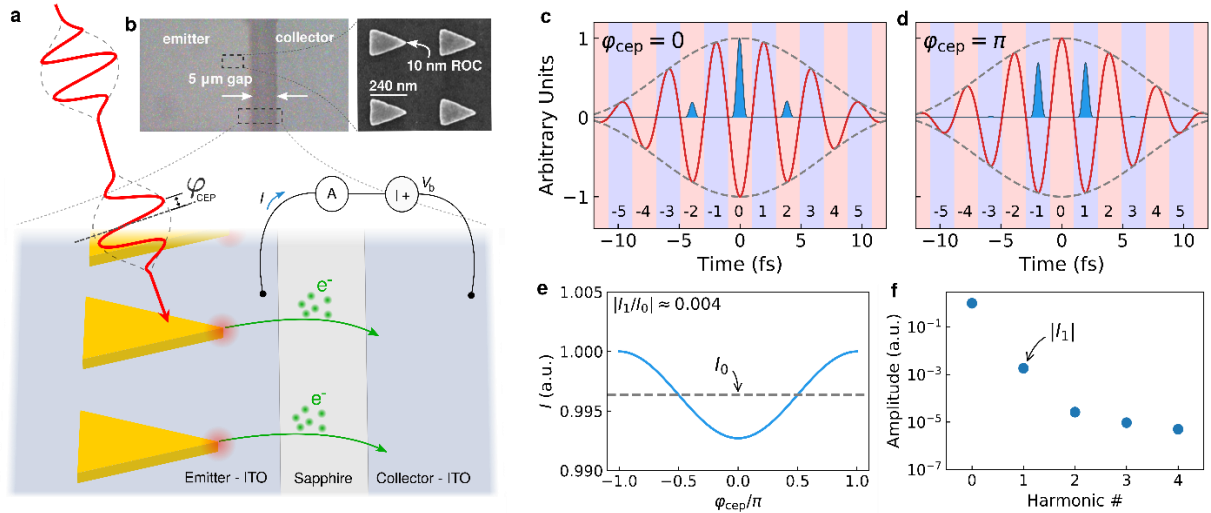

**Fig. 1** Experimental setup and illustration of CEP-sensitivity. **a.** Illustration showing optical excitation and charge extraction from the nanoantenna array in the experimental setup. Photocurrent is pulled from the nanoantenna array into an isolated collector electrode where it is then detected using a transimpedance amplifier. **b.** Optical microscope image of device geometry, including the emitter array and collector region (left) and scanning electron micrograph of triangular gold nanoantennas used in the experiment (right). **c.** Simulated waveform (red curve) along with calculated electron emission rate (filled blue curves) as a function of time for a peak field of $F_0 = 15$ GV/m with $\varphi_{cep} = 0$. Half-cycle numbers are labelled across the bottom. Note the half-cycles regions are determined relative to the center of the pulse intensity envelope, not the field waveform (*i.e.* they do not depend on $\varphi_{cep}$). **d.** Same as **c**, but with $\varphi_{cep} = \pi$. **e.** Plot of the total emitted current as a function of $\varphi_{cep}$. The average total current $I_0$ is labeled. The magnitude of the CEP-sensitivity was found to be $|I_1/I_0| = 0.004$. **f.** Magnitudes of the harmonic components of the CEP-

sensitive current $I_n$ for $n = 0$ to $n = 4$.

To understand the nature of electron emission from the nanoantennas in the optical-field emission regime, consider an optical electric field waveform at the apex of one of the nanotriangles and pointing along the surface normal. The electric field can be written as $F(t) = \Re\{F_0 \xi(t) \exp[j(\omega t + \varphi_{\text{cep}})]\}$, where $F_0$ is the peak electric field strength, $\xi(t)$ is the complex amplitude envelope, $\omega$ is the center frequency, $\varphi_{\text{cep}}$ is the carrier-envelope-phase, and $j$ is the imaginary constant. This field dynamically modulates the surface potential barrier, and if sufficiently strong (on the order of 10 GV/m or greater), the field can pull the barrier down so far as to enable a significant number of electrons to tunnel from the gold nanoantenna into the surrounding air or vacuum before it reverses polarity. In this optical-field photoemission regime, the electron emission follows a quasi-static tunneling rate[1,8,18–20]. In Figs. 1c,d we show the emission current predicted by a quasi-static Fowler-Nordheim tunneling model (shaded blue pulses) for the laser fields shown (red curves). We note that the electron current is emitted in sub-cycle bursts during half-cycles where the field is negative. Emission is suppressed during half-cycles where the field is positive as the surface electrons experience a force that drives them further into the emitter. These laser fields have a peak field strength of 15 GV/m and two different carrier envelope phase values: $\varphi_{\text{cep}} = 0$ and $\varphi_{\text{cep}} = \pi$, for Fig.1c and Fig. 1d respectively. As it will be convenient for later analysis, in Figs. 1c,d we have shaded and numbered half-cycle regions of the pulse (even half-cycles are shaded red, odd blue). We emphasize that for our purposes these half-cycle regions are defined relative to the center of the intensity envelope, not the underlying field waveform, and thus do not move when the CEP is shifted.

To quantify the impact of $\varphi_{\text{cep}}$ on the emitted current, we first need to define the key observables for both our measurements and subsequent analysis. Note that the value of $\varphi_{\text{cep}}$ strongly impacts the emission profile in time; as $\varphi_{\text{cep}}$ is changed from 0 (Fig. 1c) to $\pi$ (Fig. 1d), the total emitted charge per pulse switches from being dominated by even half-cycle contributions (Fig. 1c) to being dominated by odd half-cycle contributions (Fig. 1d). Integrating the electron emission profile over time, we find the emitted charge per unit area which is proportional to the total photoemission current from a single nanoantenna (the constant of proportionality is defined by the repetition rate of the laser pulse train and effective surface area of the emitter tip). Therefore, integrating quasi-static emission currents calculated for pulses of varying CEP, we can estimate the CEP-dependent behavior of the total photoemission current which we denote as $I_{\text{tot}}(\varphi_{\text{cep}})$. The results of this calculation are displayed in Fig. 1e (the total photoemission current has been normalized). Furthermore, performing harmonic analysis, the total photoemission current can be written as $I_{\text{tot}}(\varphi_{\text{cep}}) = \sum_n |I_n| \cos(n\varphi_{\text{cep}} + \angle I_n)$ where $I_n$ is the complex amplitude of the $n^{\text{th}}$ harmonic of the CEP-dependent total photoemission current. In Fig. 1f we plot $|I_n|$ for $n = 0$ to $n = 4$ on a logarithmic scale. We should note that $I_{\text{tot}}$ is dominated by the CEP-independent average total photocurrent $I_0$ and the first harmonic, $I_1$, *i.e.* the fundamental CEP-dependent sinusoidal component of the photocurrent. We also define the complex ratio $I_1/I_0$ as the CEP-sensitivity. In the remainder of the paper we focus on the $I_1$ component of the CEP-sensitive photoemission, but note that similar analysis could be extended to the other harmonics.

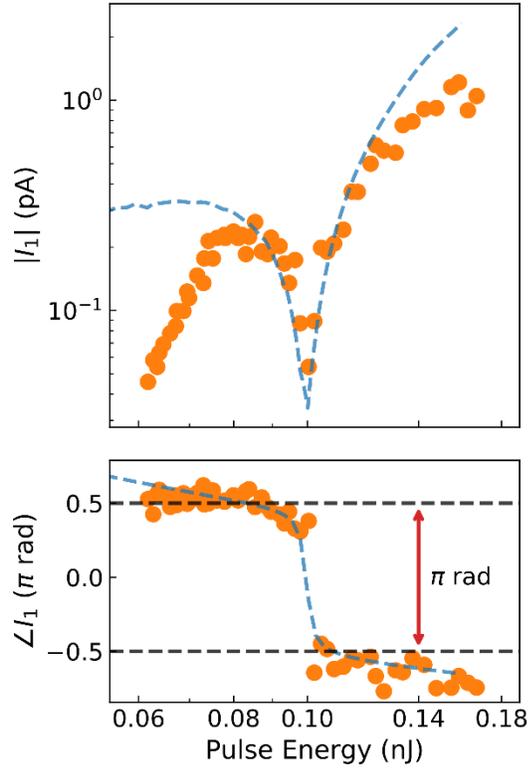

**Fig. 2** Measurement of $|I_1|$ and $\angle I_1$ as a function of the incident pulse energy (orange circles). Model results (blue dashed curve) are overlaid on top of the experimental data. At a pulse energy of $\approx 0.1$ nJ, a simultaneous dip in $|I_1|$ and phase jump of $\pi$-radians in $\angle I_1$ are observed. Our quasi-static model provides an excellent fit to the experimental data when using a field enhancement factor of $32.6 \times$, a resonant wavelength of $\lambda_{res} \approx 1105$ nm, and a damping time of $\tau = 6.8$ fs, which match well with values obtained from separate measurements.

To measure $I_1$ experimentally we locked the carrier-envelope-offset frequency $f_{ceo}$ of our incident pulse train to a stable reference at 100 Hz. In this locked state, the n[th] pulse in the train had a CEP phase of $\varphi_{cep}[n] = 2\pi n\, f_{ceo}/f_{rep} + \varphi_0$, where $f_{rep}$ is the repetition rate of the incident pulse train and $\varphi_0$ is an absolute phase offset. Using lock-in detection, we measured the magnitude and phase of the component of the photoemission current oscillating at $f_{ceo}$; this magnitude and phase correspond directly to $|I_1|$ and $\angle I_1$ respectively. Additionally, we verified that our measured $I_1$ was indeed directly controlled by the CEP of the incident pulse by translating a barium-fluoride wedge through the beam[1] (see results in Extended Data Fig. 1 and Supplementary Information Section I). Next, using a variable neutral density filter, we scanned through a range of incident pulse energies while monitoring $I_1$; the experimental results are plotted in Fig. 2 (orange circles).

Intuitively, it would seem that at higher intensities, when the overall photoemission current is larger and when we are deeper into the optical-field photoemission regime, we would observe greater CEP-sensitive photocurrent. However, we find a striking behavior in the experimental data where at a critical pulse energy of $\approx 0.1$ nJ we observe a dramatic dip in $|I_1|$ and a corresponding phase-shift of $\pi$-radians in $\angle I_1$. These behaviors are reminiscent of the characteristics of an antiresonance. (Although antiresonance-like, these behaviors appear with pulse energy, not frequency, on the horizontal axis and are not indicative of a true antiresonance in the traditional meaning of the word.) We should note that these behaviors were measured multiple times at various locations on the sample surface over several days of measurement (see Extended Data Fig. 2 and Supplementary Information Section II).

Overlaid on our experimental results in Fig. 2 are the predicted CEP-dependent behaviors from a quasi-static tunneling model (shown as a blue dashed line). The quasi-static tunneling model starts with a $\cos^2$-shaped model optical pulse. This model pulse derives from a fit to the time-domain measurement of the experimental pulse that was characterized via two-dimensional spectral shearing interferometry (2DSI)[21]. The expected field profile at the nanoantennas' apices is then approximated by filtering this model pulse with a harmonic oscillator resonator model and multiplying by a field enhancement factor. The harmonic oscillator model captures the resonant plasmonic effects at the nanoantennas[1,22,23], and the parameters for the model, namely the damping time and the resonant wavelength, follow from fits to measurements of the nanoantennas' extinction spectra (see Extended Data Fig. 3 and Supplementary Information Section III). Due to both resonant and geometric enhancement effects, there is a field-enhancement factor at the apices of the nanoantennas. This field-enhancement factor is approximated by fitting a Fowler-Nordheim (FN) emission curve to measurements of $I_0$, *i.e.* the average total photoemission current, at different intensities in the optical-field emission regime[1,20] (see Extended Data Fig. 4 and Supplementary Information Section IV). With an estimate for the field at the nanoantennas' apices, we calculated the quasi-static FN emission from our nanoantennas, and after averaging this emission over the focal spot of the incident beam, we extracted the expected CEP-sensitivity, *i.e.* $I_1/I_0$. Multiplying this sensitivity by our measurement of of $I_0$, we found the CEP-dependent current.

From Fig. 2, we see that the quasi-static model shows excellent agreement with the experimental data in both magnitude and phase, especially at high intensities where the photoemission is clearly in the optical-field-emission regime. We should note that fitting FN emission models to nine measurements of the average total photoemission current scaling versus intensity yielded field enhancement factors ranging from 28.9 × to 35.5 × with an average field enhancement of (32.6 ± 2.4) × (see Extended Data Table 1, Extended Data Fig. 4, and Supplementary Information Section IV). For the simulation results shown in Fig. 2, the field enhancement value was taken to be 32.6 × to match the average measured field enhancement, and $I_1$ was determined by multiplying the calculated CEP-sensitivity with the average total photocurrent versus intensity measurement that had a field enhancement closest to this average (for reference, the sensitivity of the model to changes in the input parameters such as the field enhancement and $\tau$ is explored in more detail in

Extended Data Fig. 5 and the Supplementary Information Section V). Fits to the nanoantennas' extinction spectra between 700-1400 nm yielded $\lambda_{res} \approx 1105$ nm and $\tau$ values ranging from 6.4 to 6.9 fs (see Extended Data Fig. 3 and Supplementary Information Section III). When modeling the antiresonance-like behavior, we used $\lambda_{res} \approx 1105$ nm and found that a damping time of $\tau = 6.8$ fs provided a fit in agreement with the experimental data. We emphasize that changes to the damping time value of less than one hundred attoseconds led to visible discrepancies between the measured and simulated antiresonance-like profiles. We note that while recent results show that tightly-focused, broadband optical pulses exhibit strong phase shifts in excess of the predicted Guoy phase near the region of the focus[24], we can safely ignore this effect here as the emitters are only 20 nm tall and were located near the minimum beam waist where the CEP does not exhibit transverse dependence. Lastly, we point out that the discrepancy between our model results and the experimental data is most pronounced at low intensities; this is not unexpected as at these intensities the emission approaches the multiphoton regime, and thus the quasi-static FN model poorly estimates the total current response.

The simulations results shown in Fig. 2 accurately account for the experimental behaviors and give us confidence that our modeling of optical-field photoemission as a quasi-static tunneling process is reasonable. However, the simulation results provide little insight into the physical origin of the antiresonance-like features in the CEP-dependent photocurrent. To get a sense for these origins, we now turn to a simplified version of the model presented in Fig. 2. In this simplified model, we ignored the filtering of the incident pulse by the nanoantennas as well as the focal spot averaging. We simply used a transform-limited pulse with similar duration and wavelength to that used in the experiment

(10 fs full width at half-maximum centered at 1177 nm) and considered emission from a single nanoantenna. The results are outlined in Fig. 3. As we discuss in detail below, we again observe antiresonance-like behavior, and we find that it arises due to competition between sub-cycle electron emission from even and odd half-cycles of the optical pulse.

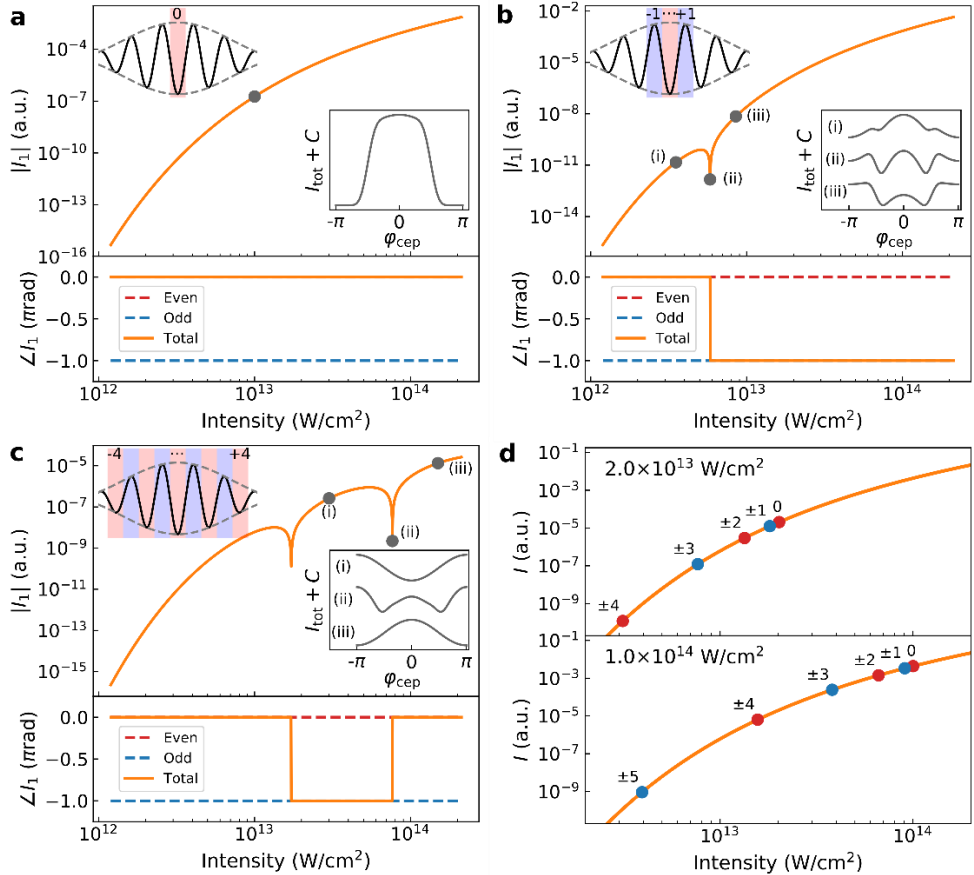

**Fig. 3** Study of antiresonance-like behavior in CEP-sensitive photocurrent excited by a transform-limited pulse. **a-c**. Plot of $|I_1|$ (top panel) and $\angle I_1$ (bottom panel) as a function of peak intensity for a transform-limited pulse with FWHM of 10 fs and central wavelength of 1177 nm when only considering emission from the shaded region shown in the top-left inset. Even half-cycles are shaded red, and odd half-cycles are shaded blue. This color scheme is used throughout. The curves in additional insets show the normalized total CEP-sensitive current response $I_{tot}$ as a function of $\varphi_{cep}$ for the intensities indicated by the gray markers (i) just before the antiresonance-like feature, (ii) at the critical intensity of the antiresonance-like feature, and (iii) just after the antiresonance-like feature. For all cases, a constant offset $C$ is added to $I_{tot}$ for clarity. The offset $C$ is different for each curve so they are separated vertically. The red and blue dashed lines in the bottom panels indicate the phase response for even half-cycles (*e.g* $0, \pm 2, \pm 4, ...$) and odd half-cycles (*e.g.* $\pm 1, \pm 3, \pm 5, ...$) respectively. **d**. Total emitted current for a single optical half-cycle as a function of intensity. Labeled markers show peak emission for each half-cycle (as labeled in Fig 1c,d) for a transform-limited pulse having a peak intensity of $2 \times 10^{13}$ W/cm² (top) and $1 \times 10^{14}$ W/cm² (bottom).

To visualize the competition between photoemission from even and odd half-cycles, in Figs. 3a-c we plot $|I_1|$ and $\angle I_1$ as a function of intensity assuming emission only from the red and blue shaded regions of the pulse shown in the top-left inset of each plot. Note that the shaded regions encompass an increasing number of half-cycles moving from the center of the pulse out to the wings: in Fig. 3a only one half-cycle is shaded, in Fig. 3b three are shaded, and in Fig. 3c nine are shaded. Also note that we enumerate the half-cycles with the zeroth half-cycle being at the center of the pulse (with $\varphi_{\text{cep}} = 0$) and positive half-cycles to the right and negative ones to the left. As with Figs. 1c,d, in Figs. 3a-c we shade even half cycles in red and odd half-cycles in blue, and this color scheme is used throughout. In the lower-right inset of the $|I_1|$ plot in Figs. 3a-c, we plot the normalized total charge emission $I_{\text{tot}}$ as a function of $\varphi_{\text{cep}}$ after being shifted by a constant $C$ for the selected intensity value(s) indicated by the gray markers. For cases where antiresonance-like features are present, such as in Figs. 3b,c, we select three intensities: (i) just before the antiresonance-like feature; (ii) at the critical intensity of the antiresonance-like feature; and (iii) just after the antiresonance-like feature. In each plot of $\angle I_1$, two reference curves called "even" and "odd" are drawn as red and blue dashed lines respectively. These reference curves indicate the phase response for even half-cycle regions (e.g $0, \pm 2, \pm 4, ...$) and odd half-cycle regions (e.g. $\pm 1, \pm 3, \pm 5, ...$) respectively. In general, the peak emission occurs near $\varphi_{\text{cep}} = 0$ for the even half-cycle regions (red), and near $\varphi_{\text{cep}} = \pm \pi$ for the odd half-cycle regions (blue). The fact that the emission dependence of even and odd half-cycle contributions are naturally $\pi$ out of phase with respect to $\varphi_{\text{cep}}$ is the root cause of the antiresonance-like behavior.

We start with Fig. 3a, where we only analyze emission from half-cycle 0. As expected, the total emitted current is maximized when $\varphi_{cep} = 0$ (tunneling requires a negative field, and we have defined $\varphi_{cep} = 0$ to be when the negative field peak is centered with the intensity envelope of the waveform). With no other competing half-cycles, this behavior holds for all intensities. In Fig. 3b we add in the $\pm 1$ half-cycles, which leads to an antiresonance-like behavior near a peak intensity of $6 \times 10^{12}$ W/cm². At a critical intensity near $6 \times 10^{12}$ W/cm² the contribution from half-cycle 0 is equal to the contribution from half-cycles $\pm 1$, which leads to a local minimum in $|I_1|$ as the two responses are out of phase, and moving past the antiresonance-like point, we observe a sudden shift of $\angle I_1$ by $\pi$-radians as the total emission switches from being dominated by electrons from half-cycle 0 to electrons from half-cycles $\pm 1$. Likewise, it is visually apparent when examining $I_{tot}$ at intensity locations (i)-(iii) that the first harmonic of the CEP-sensitive current switches from being more even-half-cycle-like with a peak near $\varphi_{cep} = 0$ to being more odd-half-cycle-like with a peak near $\varphi_{cep} = \pm\pi$. At the critical intensity indicated by (ii) the first harmonic component of the CEP-sensitive current is largely reduced due to competing emission between the 0 and $\pm 1$ half-cycles, leaving mainly the second and higher order harmonic components. As successive half-cycles are included, there continues to be competition from even and odd half-cycle contributions, until the optical field strength, and hence the emitted current, of the higher-numbered half-cycle regions weakens significantly at which point $I_1$ converges (*i.e.* $I_1$ stops changing significantly as more half-cycles are included in the calculation). For the intensity range shown, $I_1$ converges after summing over half-cycles $-4$ to $+4$ (see Fig. 3c), demonstrating that only the central-most half-cycles control the CEP-sensitive photocurrent response.

To show why there is a switch in dominance between even and odd half-cycle emission regions, we plot the emission rate predicted by the quasi-static Fowler-Nordheim model as a function of peak intensity in Fig. 3d; on top of this plot, we show markers that correspond to the peak emission rate for the labeled half-cycle regions inside of the transform-limited optical pulse (the labeled and shaded regions in Fig. 1a-c) for two different peak pulse intensities: $2 \times 10^{13}$ W/cm$^2$ (top) and $1 \times 10^{14}$ W/cm$^2$ (bottom). The fact that the pulse is transform-limited explains the degeneracy between the left- and right-half contributions to the emission, as for a chirped pulse, the emission from the left- and right-half contributions would differ. Since the rate of increase in electron emission declines at high intensities, the ratio of emission between even and odd numbered half-cycles also changes with intensity. With this in mind, consider again Fig. 3c. At the lowest intensities, emission from even half-cycle regions dominates. However, as intensity is increased we can see from Fig. 3d that the rate of increase in the emission from half-cycle 0 starts to decline, allowing the combined emission from the odd half-cycle regions to approach and eventually overcome that of the even half-cycle regions. As the contribution to $I_1$ from the even half-cycle regions is out-of-phase by $\pi$ radians compared to that of the odd half-cycle regions, there is an antiresonance-like event when the contribution to $I_1$ from even and odd half-cycle regions is equal. Eventually, as the intensity is further increased, the rate of increase in emission from half-cycles $\pm 1$ starts to decline, and the contribution to $I_1$ from the even half-cycle regions approaches and then overcomes that of the odd half-cycle regions and there is another antiresonance-like event. Thus, a critical requirement for the appearance of antiresonance-like features is a nonlinear change in emission rate as a function of intensity, which results in an intensity-dependent change in the relative contributions to total electron emission from even and odd half-cycles (see Extended Data Figs. 6-8 and Supplementary Information Section VI for discussion of antiresonance-like features arising from atomic and electronic systems).

Note that in Fig. 3a-c, with each antiresonance-like feature, the phase jumps instantaneously between 0 and $\pm\pi$. This is due to the fact that the pulse is transform-limited and perfectly symmetric in time. However, in the experimental and simulation results which use realistic pulses with residual chirp and plasmonic damping, we note that the phase response of the antiresonance-like feature is smoother with a slight downward slope. This is due to the asymmetry of the optical pulse in time, which causes $\angle I_1$ to adiabatically shift with increasing intensity. The greater the chirp or asymmetry in the pulse shape, the more pronounced this slope in $\angle I_1$ as a function of incident intensity. By examining the antiresonance-like response in detail, including the slope of $\angle I_1$ and the precise location and depth of the dip in $|I_1|$, one can likely obtain a great deal of information about the time-domain profile of the optical waveform at the surface of the emitter. (See sensitivity of antiresonance-like features to changes in pulse dispersion in Extended Data Fig. 9 and the Supplementary Information Section VII.)

As a last note, we point out that these antiresonance-like features arise purely from the field-dependent nature of quasi-static tunneling, and the fact that the tunneling emission growth rate decreases as field strength increases. This makes the appearance of antiresonance-like features in the CEP-sensitive current response a general phenomenon that should be observable for any system where the current is field driven with similar nonlinear characteristics to that of Fowler-Nordheim-like tunneling. For instance, we can reproduce this behavior using models of CEP-sensitive current arising from few-cycle optical-field emission from atoms, as well as CEP-sensitive current arising from few-cycle voltage pulses applied to a diode in series with a resistor (see Extended Data Fig. 6 and Supplementary Information Section VI). This makes such antiresonance-like features ideal candidates for carefully examining the ultrafast emission response from any system where nonlinear, field-driven currents dominate.

In summary, we have experimentally and theoretically characterized an antiresonance-like phenomenon in the intensity-scaling of CEP-sensitive photocurrent from plasmonic nanoantennas. We found this antiresonance-like behavior to arise from competing sub-optical-cycle, sub-femtosecond bursts of charge. The sensitivity of the antiresonance-like behavior to subtle changes in the optical electric field allowed us to simultaneously confirm both the field enhancement and the damping time of the nanoantennas' plasmonic response, which together with the central wavelength of the plasmonic resonance describe the excited optical electric field at the emitter surface. In particular, we found the antiresonance-like features to be sensitive to damping time changes of less than one hundred attoseconds. As the antiresonance-like behavior we have observed is a general consequence of tunneling emission, it should be observable in any system in the optical-field emission regime. We emphasize that the observation of antiresonance-like behavior requires harmonic analysis of the CEP-sensitive photocurrent, thus the unique way in which the CEP-sensitive current was detected and analyzed in this work using a slowly oscillating $\varphi_{\mathrm{cep}}$ and lock-in detection of only the fundamental harmonic of the CEP-sensitive current response, was of critical importance. However, this antiresonance-like behavior is not isolated to just $I_1$, as our simulations show that higher-order harmonics also exhibit antiresonance-like features (see Extended Data Fig. 6b). Given that the antiresonance-like features discussed in this work are sensitive to changes in the emission rate as well as the incident optical waveform, a similar experimental analysis could be extended to further probe the physics of strong-field electron emission dynamics in solids as well as gases, to characterize the precise shape of the optical waveform driving optical-field photoemission, or to sense subtle changes in the electronic structure of surfaces.

## Acknowledgements

We would like to thank Jim Daley for his assistance in device fabrication. This work was supported by the United States Air Force Office of Scientific Research (AFOSR) through grant FA9550-12-1-0499, the Center for Free-Electron Laser Science at DESY, and The Hamburg Center for Ultrafast Imaging: Structure, Dynamics and Control of Matter at the Atomic Scale, an excellence cluster of the Deutsche Forschungsgemeinschaft. R.G.H. acknowledges support for the device fabrication work from the Center for Excitonics, an Energy Frontier Research Center funded by the US Department of Energy, Office of Science, Office of Basic Energy Sciences under award number DE-SC0001088.

## Author Contributions

P.D.K. and W.P.P. conceived of the experiment and contributed equally to this work. W.P.P., R.G.H., and Y.Y. fabricated the devices. P.D.K., W.P.P., and P.V. performed the measurements, collected the experimental data, and performed the numerical analysis. P.D.K., W.P.P., and P.V. composed the manuscript. P.D.K., W.P.P., P.V., R.G.H., Y.Y., K.K.B., and F.X.K. interpreted the results and contributed to the final manuscript.

## Competing Financial Interests

The authors declare no competing financial interests.

## Methods

Methods used for optical pulse generation, $\varphi_{cep}$ stabilization, and nanoantenna array fabrication have been thoroughly detailed in the Supplementary Information of Ref.[1]

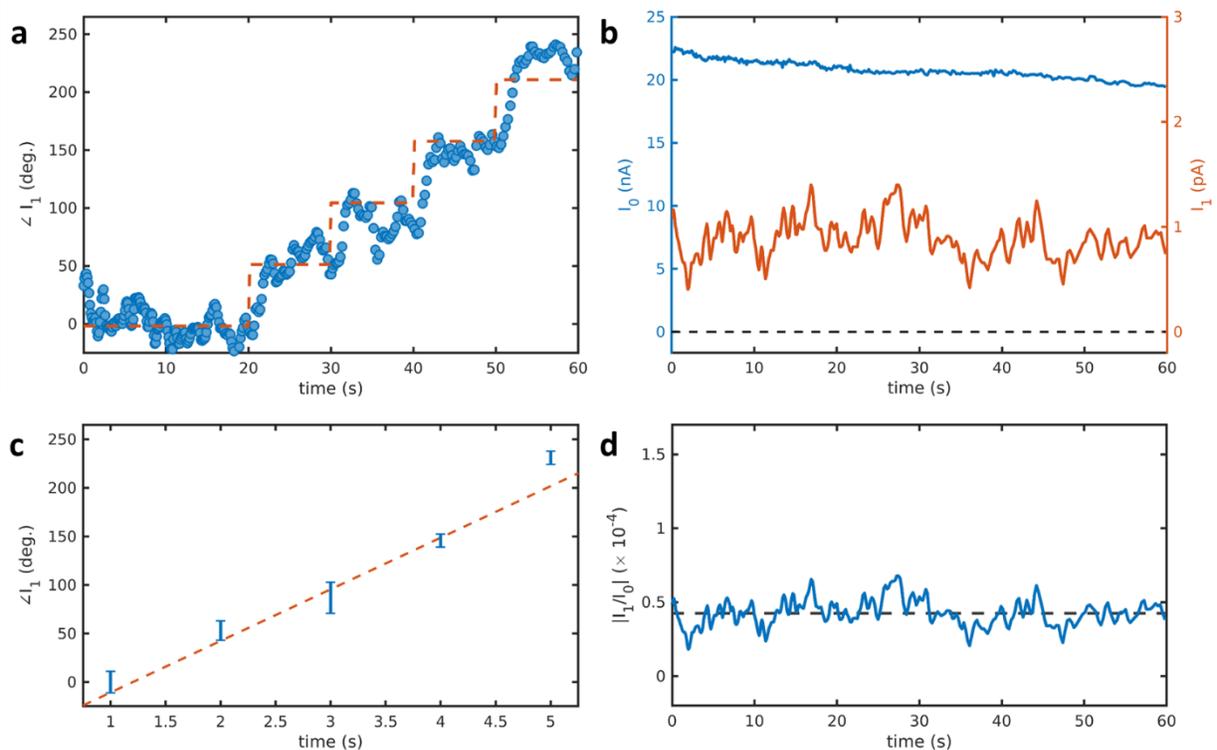

**Extended Data Fig. 1** Verification of CEP-sensitivity. **a.** During a one-minute scan, a $BaF_2$ wedge is inserted into the beam path by 2.5 mm every 10 seconds (first insertion at $t = 20$ sec). The red dashed line shows a fit to the phase stepping, with an average step size of 53.1°, which matches well to the calculated expected step size of 58.3° based on the optical properties and shape of the $BaF_2$ wedge. **b.** During the scan, $I_0$ and $I_1$ were simultaneously monitored, showing that while the wedge shifts $\angle I_1$ it has no effect on $I_1$ as one would expect. **c.** The measured phase shift per wedge insertion with the fit indicating 53.1° per step. **d.** The CEP-sensitivity magnitude, $|I_1/I_0|$, measured over the scan.

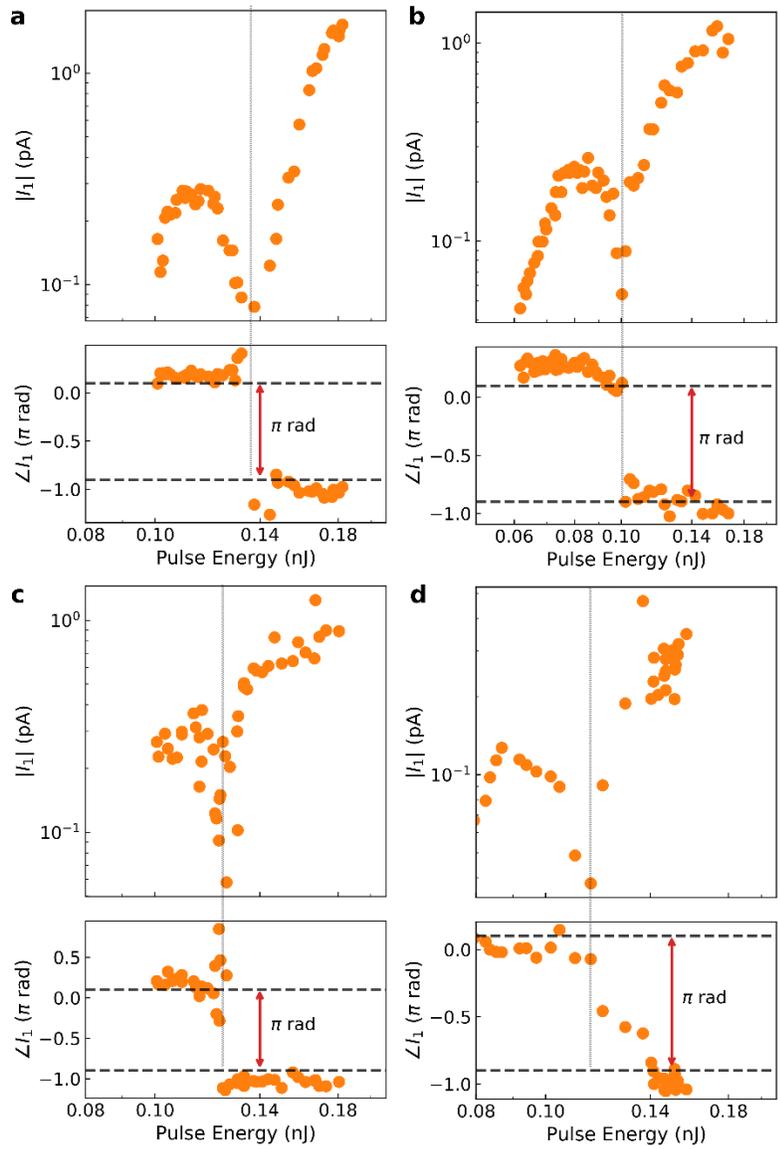

**Extended Data Fig. 2** Four separate intensity scans showing antiresonance-like behavior (**a-d**). For each scan we plot $|I_1|$ in the top plot and $\angle I_1$ in the bottom plot. For each scan, there is a sudden drop in $|I_1|$ that is accompanied by a $\pi$ phase-shift in $\angle I_1$. We note that the exact locations and widths of the dips show some variation from scan to scan.

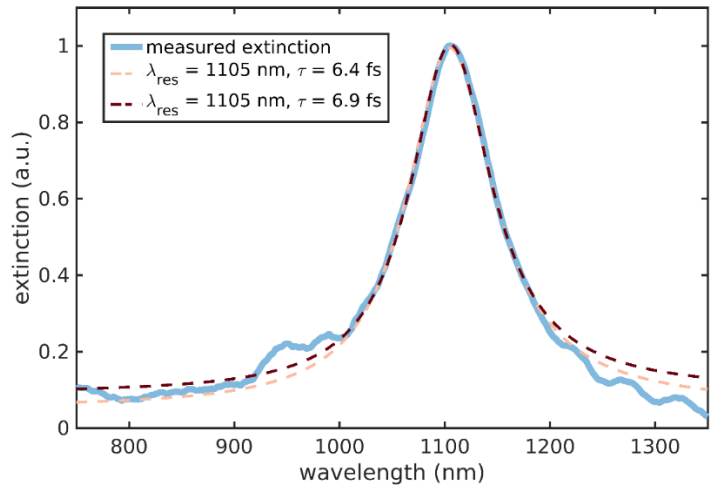

**Extended Data Fig. 3** Extinction spectra fits. Note that for the entire wavelength range spanning from 700-1400 nm, a damping time of $\tau = 6.4$ provides the best fit. However, when the wavelength range is restricted to be from 950-1250 nm, closer to the center of the resonance, the best fit is given by $\tau = 6.9$ fs. Furthermore, we note that in the center of the resonance, these two damping times lead to a negligible change in the extinction spectrum curve. For both cases, the central wavelength was found to be $\lambda_{res} \approx$ 1105 nm.

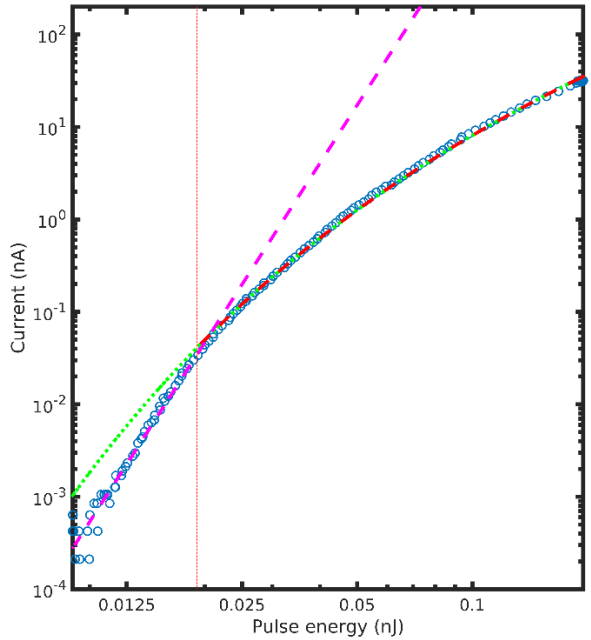

**Extended Data Fig. 4** Fitting field enhancement using the average total photocurrent. The measured average total photocurrent (blue circles) is fit using two models. A FN model (green and red dashed lines) provides a nice fit for $\gamma < 1$, which is indicated by the vertical dashed line. For $\gamma > 1$, the data follows a multiphoton emission curve (magenta dashed line) which is proportional to $P^N$, where $P$ is the pulse energy. For this measurement, we found $N \approx 6.5$ and a field enhancement of $32.1 \times$.

| Run No. | Field Enhancement |
|---|---|
| 1 | 34.1 |
| 2 | 30.7 |
| 3 | 32.1 |
| 4 | 28.9 |
| 5 | 30.5 |
| 6 | 37.3 |
| 7 | 33.7 |
| 8 | 35.4 |
| 9 | 35.5 |

**Extended Data Table 1** Field enhancement fits for nine separate current scaling runs.

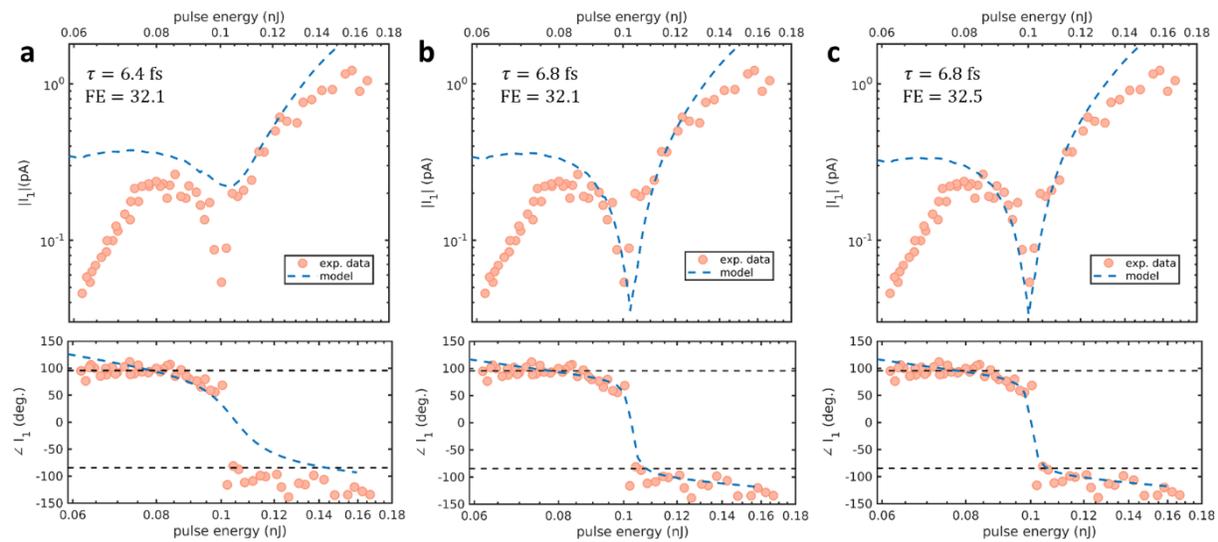

**Extended Data Fig. 5** Sensitivity of modeled $I_1$ response to $\tau$ and $FE$. Changes in $\tau$ tend to modify the depth and width of the dip, while changes in $FE$ adjust the location of the dip. **a.** Model results compared to experimental data for $\tau = 6.4$ fs and $FE = 32.1$. **b.** For $\tau = 6.8$ fs and $FE = 32.1$. **c.** For $\tau = 6.8$ fs and $FE = 32.5$.

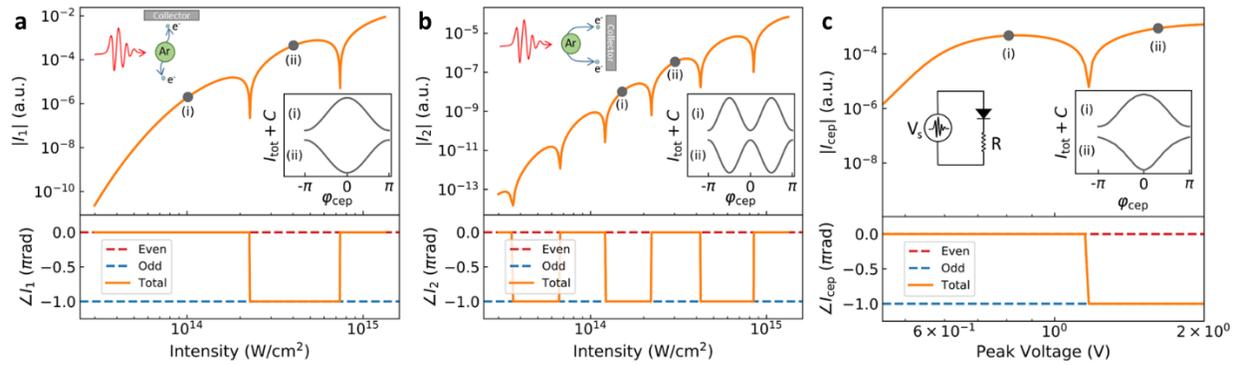

**Extended Data Fig. 6** Antiresonance-like behavior in atomic and electronic systems. (a) Antiresonance-like behavior in calculated single-sided emission from an argon atom. The calculation was based on the ADK tunneling rate under the single-active electron approximation, using an ionization potential of 15.76 eV. (b) Antiresonance-like behavior in calculated double-sided emission from an argon atom. (c) Antiresonance-like behavior in calculated current from a diode in series with a resistor. The diode turn-on voltage was taken to be 25 mV, and the resistance value was chosen to be 10 Ω. The voltage pulse was roughly two cycles FWHM.

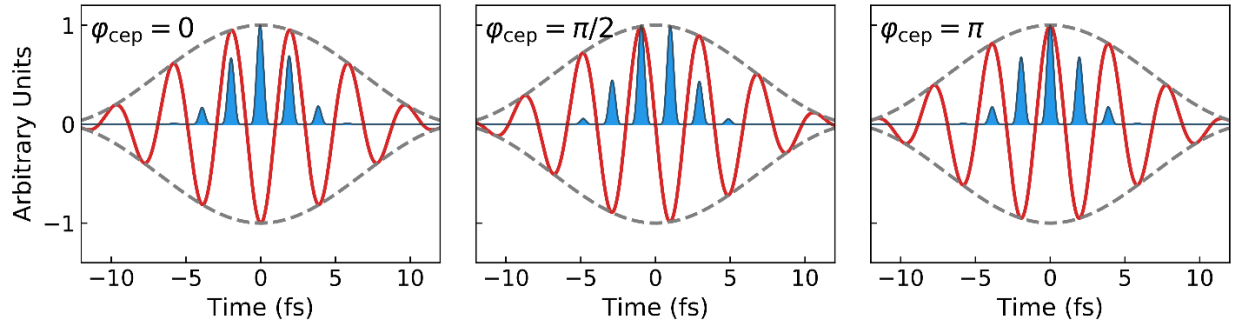

**Extended Data Fig. 7** Calculated double-sided emission from an argon atom for peak field strength of 50 GV/m at three $\varphi_{cep}$ values of the optical pulse. Due to the inversion symmetry, there is equal amount of emission for positive and negative fields of the same strength. This leads to a $\pi$-periodicity with respect to $\varphi_{cep}$ for double-sided emission.

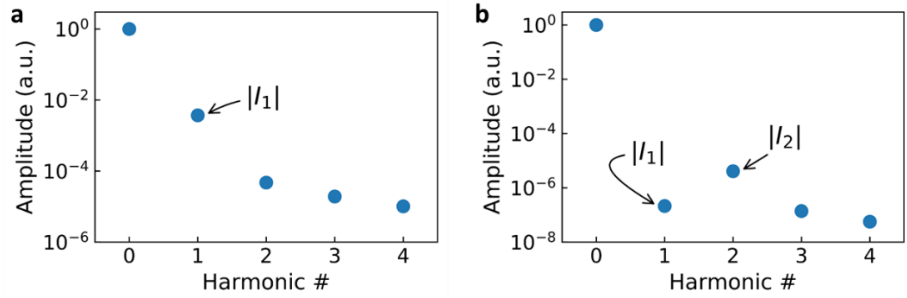

**Extended Data Fig. 8** Comparison of the harmonic analysis of $I_{tot}$ for a peak field strength of 50 GV/m for the case of (**a**) single-sided emission from argon, and (**b**) double-sided emission from argon. Note that the average total current $I_0$ has been normalized to 1 in both cases. For the case of double-sided emission, $I_1$ is suppressed significantly due to the inversion symmetry.

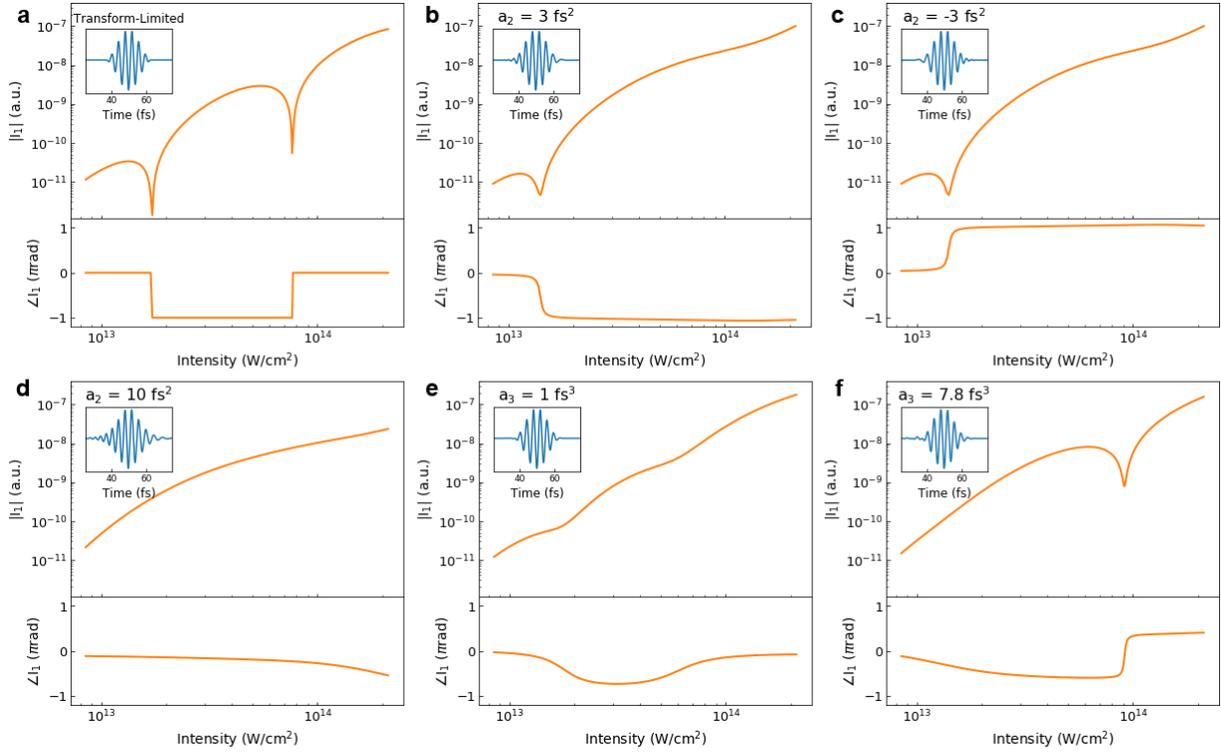

**Extended Data Fig. 9** Sensitivity of antiresonane-like behavior to pulse shape. **a** Scaling of CEP-dependent current magnitude and phase with intensity for the transform-limited optical pulse. **b** $a_2 = 3$ fs$^2$ second-order dispersion term is added to the transform-limited pulse. We note a shift in the intensity value of the dip in magnitude, a smoother phase transition, and only a single antiresonance-like switch over the same intensity range. **c** $a_2 = -3$ fs$^2$ second-order dispersion term is added to the transform-limited pulse. Compared to $a_2 = 3$ fs$^2$, we see a reflection in phase scaling with intensity, a property that is observed for sign changes in both $a_2$ and $a_3$. **d** $a_2 = 10$ fs$^2$ second-order dispersion term is added to the transform limited pulse. We find that the antiresonance-like effect disappears entirely. **e** $a_3 = 1$ fs$^3$ third-order dispersion term is added to the transform-limited pulse. We find less dramatic dips in magnitude and smoother phase transitions. **f** $a_3 = 7.8$ fs$^3$ third-order dispersion term is added to the transform-limited pulse. This value demonstrates that for various values of $a_3$, we reintroduce the antiresonance-like effect.

# Supplementary information

## I. CEP Sensitivity

The CEP-sensitivity was verified by inserting a BaF$_2$ wedge into the beam while monitoring $\angle I_1$ using the lock-in amplifier. The CEP-phase due to the wedge insertion was calculated and compared to the measured change in $\angle I_1$. An example of such a scan is shown in Extended Data Fig. 1.

## II. Repeatability of the Antiresonance-Like Behavior

The antiresonance-like switching behavior was observed during multiple scans over three days of testing. In Extended Data Fig. 2, we show four compiled scans that demonstrate antiresonance-like behavior in the CEP-sensitive photocurrent. We measure dip locations ranging from approximately 0.1 to 0.14 nJ, corresponding to field enhancement factors ranging from roughly 27.5 to 32.5. This enhancement is in-line with the field enhancement factors found using intensity scans of the total current and fitting to the Fowler-Nordheim emission rate, which were found to range between 28.9 and 35.5 (see Supplementary Information Section IV). Modeling shows that the width of the dip is related to both intensity averaging over the focal spot and the damping factor. We attribute the observed changes in the antiresonance-like dip to natural variations in the nanoantennas over various spots within the fabricated array.

## III. Extinction Spectra Fits

As described in the main text, extinction spectra of the nanoantennas were measured and fit using a damped harmonic oscillator model. The fit parameters were the central wavelength, $\lambda_{res}$, and the damping time $\tau$ of the resonance. While $\lambda_{res}$ was found to be $\approx 1105$ nm in all cases, we found that the fit was less sensitive to the precise value of $\tau$, and the fit result varied depending on the exact wavelength range used. For instance, fitting over all measured wavelengths between 700-1400 nm, we find $\tau \approx 6.4$ fs, while fitting between 950-1250 nm we find $\tau \approx 6.9$ fs. The modeled extinction curves for each case are plotted along with the measured extinction curve in Extended Data Fig. 3. Based on this analysis, we find the damping time to likely fall within the range of $\tau = 6.4$ fs to $\tau = 6.9$ fs, which is supported by our modeling of the antiresonance-like behavior in the main text, where the best fit was found for $\tau = 6.8$ fs.

## IV. Field Enhancement Determination

The field enhancement factor of the devices was determined by measuring the average total photocurrent as a function of incident pulse energy. Using the measured spot size, the incident intensity before enhancement was determined; then fitting the average total photocurrent to a Fowler-Nordheim (FN) tunneling model, the field-enhancement factor was determined. As optical-tunneling only applies when the Keldysh parameter $\gamma$ is less than one, this was an iterative process: (1) the FN model was first used for the entire dataset to obtain a first estimate; (2) only those intensity values such that $\gamma < 1$ were used to obtain the next field enhancement factor; (3) step (2) was repeated until the process converged to the final field enhancement factor value[1]. In Extended Data Fig. 4, we show the fit results for a single scan. In Extended Data Table 1, we compile field enhancement factors from 9 separate scans for reference.

## V. Sensitivity to Field Enhancement and Damping Time Parameters

The plots shown in Extended Data Fig. 5 demonstrate the effects of different field enhancement and damping time values on the calculated CEP-sensitive current yield. We compare the calculations to the experimental data used in the article for reference.

## VI.     Antiresonance-Like Behavior in Other Systems

As mentioned in the main text, the antiresonance-like behavior we observed in CEP-sensitive photoemission from nanostructures arises from general properties of field-driven emission. Our understanding thus far is that the critical requirements for observing antiresonance-like behavior in the CEP-sensitive emission are that the emission be driven directly by the field, and that emission rate changes nonlinearly as a function of intensity. This is true for atomic systems as well as metallic surfaces, which we demonstrate by using the Ammosov, Delone, Krainov (ADK) tunneling rate[25] to calculate $I_1$ arising from an argon atom under the single-active electron approximation. One complication that arises from atomic systems, or any system with inversion symmetry, such as nanorods[1], is that the emission can arise from either side of the atom relative to the polarized optical electric field. We start by assuming that only electrons arising from one side of the atom are collected (see top left inset of Extended Data Fig. 6a). The results of this calculation are shown in Extended Data Fig. 6a.

Note the similarity of the results in Extended Data Fig. 6a to those presented in Fig. 3 of the letter. The only obvious difference is the increase in the relative intensity values due to the larger ionization potential (the ionization potential of argon is 15.76 eV versus a work function of only 5.1 eV for gold). Interestingly, if electrons from both sides of the atom are collected, the same antiresonance-like behavior is observed, only now, due to the inversion symmetry, there is a $\pi$-periodicity relative to $\varphi_{\text{cep}}$ as shown in Extended Data Fig. 7. This $\pi$-periodicity suppresses $I_1$ meaning that the dominant CEP-sensitive signature becomes $I_2$ as shown in Extended Data Fig. 8[1]. Nevertheless, the same antiresonance-like behavior arises, only now more frequently, with the total CEP-sensitive current now switching between even and odd quarter-cycle dominance as shown in Extended Data Fig. 6b.

This antiresonance-like behavior is not restricted to optical systems. The criteria for antiresonance-like behavior is also satisfied by few-cycle voltage pulses being applied to a diode in series with a resistor. As a final example, we perform such a calculation, modeling a roughly two-cycle full-width at half-maximum (FWHM) voltage pulse driving a diode with a turn-on voltage of 25 mV in series with a 10 Ω resistor. The results are shown in Extended Data Fig. 6c. A clear antiresonance-like feature appears as the diode current becomes limited by the resistor. Carrying the concept further, the double-sided emission analogy could even be reproduced using a bridge-rectifier (not shown).

## VII. Sensitivity to Optical Pulse Shape

As evidenced by its sensitivity to small changes in the damping time of the plasmonic resonator, we find in general that the antiresonance-like behavior is highly sensitive to the precise shape of the optical waveform in the time domain. To examine this in further detail, we used our quasi-static FN emission model to simulate the emission from a $\cos^2$-envelope optical pulse with a transform-limited duration of 10 fs FWHM, a central wavelength of 1177 nm, and varying degrees of second- and third-order dispersion. The dispersion was added to the transform-limited pulse in the frequency domain, being $\mathcal{F}_{TL}(\omega)$, with the stretched pulse written as $\mathcal{F}_{stretched}(\omega) = \mathcal{F}_{TL}(\omega) \exp(j[\varphi_{cep} + a_1(\omega - \omega_0) + a_2(\omega - \omega_0)^2 + a_3(\omega - \omega_0)^3])$ where the $a_i$ are real constants (with units of fs$^i$) that were varied manually and $\omega_0$ the central frequency of the pulse in fs$^{-1}$. The time-domain form of the chirped pulse was then taken to be the inverse Fourier transform of $\mathcal{F}_\text{stretched}(\omega)$. We note that $a_1$ only varies the location in time of the pulse so we did not study its effect here as it would have no impact on the CEP-sensitivity of the emission.

We find that this antiresonant-like behavior is highly sensitive to changes in $a_2$ and $a_3$. Key results of our analysis on the impacts of $a_2$ and $a_3$ are shown in Extended Data Fig. 9. In Extended Data Fig. 9a, we first show the intensity scaling of $|I_1|$ and $\angle I_1$ for the transform-limited pulse. We find the two antiresonance-like switches seen previously in Fig. 3c. In Extended Data Fig. 9b, we add second-order dispersion to the pulse and find that the intensity value corresponding to the leftmost switch in Extended Data Fig. 9a changes as does the depth of the dip in $|I_1|$. Additionally, we see that the rightmost switch found in Extended Data Fig. 9a disappears entirely. In Extended Data Fig. 9c, we flip the sign of the second-order dispersion used in Extended Data Fig. 9b. We see that the intensity scaling of $|I_1|$ is identical between Extended Data Figs. 9b-c but $\angle I_1$ shifts in opposite directions. This reflection of $\angle I_1$ depending on the sign of the chirp parameter is seen for variations in both $a_2$ and $a_3$. In Extended Data Fig. 9d, we apply a larger amount of second-order dispersion to the pulse and find that the antiresonance-like response disappears entirely. The value of $\angle I_1$ remains nearly constant as intensity is increased, and no drop-off in $|I_1|$ is observed.

In Extended Data Fig. 9e, we now add a small amount of third-order dispersion and observe that $|I_1|$ does not exhibit sharp dips but instead smoothly decreases at intensity values corresponding to the switching intensities of Extended Data Fig. 9a. We also see that $\angle I_1$ shifts by less than $\pi$-radians at both of these intensity values, leading to a shallower, smoother version of the $\angle I_1$ plot of Extended Data Fig. 9a. In Extended Data Fig. 9f, we demonstrate that for many different values of the third-order dispersion parameter $a_3$, we are able to reintroduce antiresonance-like behavior to the intensity scaling. The shown value of $a_3 = 7.8$ fs$^3$ was arbitrarily found- we find many such values of $a_3$ for which antiresonance-like behavior is exhibited.

Using these results, we show that the intensity scaling of $|I_1|$ and $\angle I_1$ contains information about the properties of the pulse in the time domain. This motivates the possibility of partially or fully characterizing the driving pulse *in situ* by studying the intensity scaling of $|I_1|$ and $\angle I_1$. This is important as current techniques characterize the ultrafast plasmonic response of the nanoantenna in an indirect fashion. For instance, extinction spectra are analyzed to develop a model of the plasmonic response, and the field at the emitter tip is arrived at by filtering a characterization of the optical pulse before interaction.